\documentclass[fleqn,usenatbib]{mnras}
\usepackage{newtxtext,newtxmath}

\usepackage[T1]{fontenc}
\usepackage{ae,aecompl}
\usepackage{graphicx}	
\usepackage{amsmath}	
\usepackage{amssymb}	

\usepackage{mathtools}
\usepackage{cases}
\usepackage{cuted}
\usepackage{bigstrut}
\usepackage{subfig}
\usepackage{paralist}
\usepackage{lscape}
\usepackage{wrapfig}
\usepackage[T1]{fontenc}
\usepackage{aecompl}
\usepackage{adjustbox} 
\usepackage{underscore}
\usepackage{float}
\usepackage{amsfonts}
\usepackage{color}
\usepackage{booktabs}
\usepackage[flushleft]{threeparttable}
\usepackage[english]{babel}
\usepackage{lastpage}
\usepackage[compact]{titlesec}
\titlespacing{\section}{0pt}{2ex}{1ex}
\titlespacing{\subsection}{0pt}{2ex}{1ex}
\titlespacing{\subsubsection}{0pt}{2ex}{1ex}

\newcommand{\civ}{\mbox{C\,{\sc iv}}}

\newcommand{\feii}{\mbox{Fe\,{\sc ii}}}
\newcommand{\feiii}{\mbox{Fe\,{\sc iii}}}

\newcommand{\siv}{\mbox{S\,{\sc iv}}}

\newcommand{\nv}{\mbox{N\,{\sc v}}}
\newcommand{\al}{\mbox{Al\,{\sc iii}}}
\newcommand{\alii}{\mbox{Al\,{\sc ii}}}
\newcommand{\mgii}{\ifmmode {\rm Mg}{\textsc{ii}} \else Mg\,{\sc ii}\fi}
\newcommand{\heii}{\ifmmode {\rm He}{\textsc{ii}} \else He\,{\sc ii}\fi}

\newcommand{\asca}{{\it ASCA\/}}
\newcommand{\chandra}{{\it Chandra\/}}
\newcommand{\gemini}{{\it Gemini\/}}
\def\aox{$\alpha_{\rm ox}$}
\newcommand{\daox}{$\Delta\alpha_{\rm ox}$}
\newcommand{\xray}{X-ray}

\newcommand{\angstrom}{\text{\normalfont\AA}}

\setcitestyle{notesep={; },round,aysep={},yysep={,}}

\title[Investigations of transforming BAL quasars]{X-ray and multi-epoch optical/UV investigations of BAL to non-BAL quasar transformations}

\author[Sameer et al.]
{Sameer,$^{1}$\thanks{E-mail: sxx15@psu.edu}  
W. N. Brandt,$^{1,2,3}$ 
S. Anderson,$^{4}$
P. B. Hall,$^{5}$
M. Vivek,$^{1}$
N. Filiz Ak,$^{6,7}$
\newauthor
C. J. Grier,$^{1,2}$
N. S. Ahmed,$^{5}$
B. Luo,$^{8}$
A. D. Myers,$^{9}$
P. Rodr\'{\i}guez Hidalgo,$^{10}$
J. Ruan,$^{11}$
\newauthor
and D. P. Schneider$^{1,2}$
\\
$^{1}$Department of Astronomy \& Astrophysics, 525 Davey Lab,
The Pennsylvania State University, University Park, PA 16802, USA\\
$^{2}$Institute for Gravitation and the Cosmos,
The Pennsylvania State University, University Park, PA 16802, USA\\
$^{3}$Department of Physics, 104 Davey Lab,
The Pennsylvania State University, University Park, PA 16802, USA\\
$^{4}$Department of Astronomy, C319, University of Washington, Seattle WA 98195-0002, USA\\
$^{5}$Department of Physics \& Astronomy, York University, 4700 Keele Street, Toronto, ON M3J 1P3, Canada\\
$^{6}$Department of Astronomy and Space Sciences, Erciyes University, Kayseri, 38039\\
$^{7}$Astronomy and Space Sciences Observatory and Research Center, Erciyes University, Kayseri, Turkey\\
$^{8}$School of Astronomy and Space Science, Nanjing
University, Nanjing, Jiangsu 210093, China\\
$^{9}$Department of Physics \& Astronomy, University of Wyoming, 1000 E. University, Dept. 3905, Laramie, WY 82071, USA\\
$^{10}$Department of Physics \& Astronomy, Humboldt State University, 1 Harpst Street, Arcata, CA 95521\\
$^{11}$McGill University Department of Physics, 3550 Rue University, Montreal, QC, Canada H3A 2T8\\
}

\date{Accepted 2018 October 4. Received 2018 September 13; in original form 2018 June 20}

\pubyear{2018}

\begin{document}
\label{firstpage}
\maketitle

\begin{abstract}

\noindent We report on an \hbox{X-ray} and optical/UV study of eight Broad Absorption Line (BAL) to non-BAL transforming quasars at $z\,\approx\,1.7$--2.2 over 0.29--4.95 rest-frame years with at least three spectroscopic epochs for each quasar from the SDSS, BOSS, \hbox{\gemini}, and ARC 3.5-m telescopes. New \hbox{\chandra}~observations obtained for these objects show their values of \aox~and \daox, as well as their spectral energy distributions, are consistent with those of non-BAL quasars. Moreover, our targets have \xray\ spectral shapes that are, on average, consistent with weakened absorption with an effective power-law photon index of $\Gamma_{\rm eff}\,=\,1.69^{+0.25}_{-0.25}$. The newer \hbox{\gemini}~and ARC 3.5-m spectra reveal that the BAL troughs have remained absent since the BOSS observations where the BAL disappearance was discovered. The \hbox{X-ray} and optical/UV results in tandem are consistent with at least the \hbox{X-ray} absorbing material moving out of the line-of-sight, leaving an \hbox{X-ray} unabsorbed non-BAL quasar. The UV absorber might have become more highly ionized (in a shielding-gas scenario) or also moved out of the line-of-sight (in a wind-clumping scenario).

\end{abstract}



\renewcommand*{\thefootnote}{\textsuperscript{\arabic{footnote}}}

\section{INTRODUCTION}
\label{sec:Intro}

\noindent Outflows from Active Galactic Nuclei (AGN) are an integral part of the environment around the central engine driving the activity. Their near universality in AGNs indicates that mass ejection is associated with or even essential for accretion onto the central supermassive black hole (SMBH). Additionally, another critical role played by the outflows is the feedback process of SMBHs into typical massive galaxies, which could lead to quenching of the star formation and growth of the SMBH, perhaps accounting for the observed relationships between the SMBH mass and galaxy bulge properties~(e.g. \citealt{king2015powerful} and references therein). 

A robust indicator of outflows in luminous quasars is the presence of UV Broad Absorption Lines (BALs). BALs are seen as blue-shifted relative to the corresponding emission lines and have velocity widths in excess of \hbox{$2000$~km~s$^{-1}$} at absorption depths at least ten per cent below the continuum (e.g. \citealt{wey91}); their strength is typically measured with the ``balnicity index'' parameter (BI, see Equation~\ref{eqn:BI}). These features are considered to be generated by high-velocity outflowing winds launched close to the quasar's SMBH. BALs are not ubiquitous in the spectra of quasars; however, they are seen in $\approx$\,10--20 per cent of the optically selected quasar population within the redshift range \hbox{1.5\,$\leq\,z\,\leq$\,2.5} (e.g. \citealt{trump2006catalog,gibson2009catalog,allen2010strong}). BAL quasars are classified into three groups based on the observed transitions: High-ionization BAL quasars (HiBALs) show only resonance transitions such as \civ, \siv, and \nv; low-ionization BAL quasars (LoBALs) exhibit lower-ionization lines such as \alii, \al, and \mgii~in their spectra in addition to the HiBAL features; iron low-ionization BAL quasars (FeLoBALs) are LoBALs that also possess \feii~and/or \feiii~absorption lines (e.g. \citealt{hall2002unusual}). 

The main driver of BAL outflows is believed to be radiation pressure on the UV lines (e.g. \citealt{murray1995accretion,proga2000dynamics}) with a possible additional contribution from magnetohydrodynamical forces (e.g. \citealt{konigl1994disk}).
A better understanding of the physical attributes such as the structure and location of the outflows would serve to ascertain if outflows indeed are a feasible mechanism to induce feedback in their host galaxies. BAL variability serves as one handle to infer the properties of these outflows. BALs can vary in strength, velocity shift, and profile shape. The causes for variability are severalfold, e.g., variations in the amount of radiation received by the gas which could be due to intrinsic variations of the source or due to ``shielding-gas'' variations \citep[e.g.][]{ak2012,ak2013broad}, or bulk motion of the outflowing gas (e.g. \citealt{lundgren2007broad,gibson2008optically,hall2011implications,vivek2012dynamically,capellupo2013variability}).

BAL variability spans a broad temporal scale, ranging from timescales of years (e.g. \citealt{ak2013broad}) to as little as $\approx$\,1.2 days (e.g. \citealt{grier15}). The short-term variability places constraints on the absorbing material's distance from the central SMBH, while studies on long-term timescales (e.g. \citealt{ak2013broad}) have shown that \civ~BAL variability is found across a broad range of outflow velocities, variability of multiple troughs of the same ion frequently happen in a coordinated manner, and importantly, the Equivalent Width (EW) and fractional EW variations of \civ~and \siv~BAL troughs increase with sampled rest-frame time-scale. 

An equatorial accretion-disk wind model is frequently invoked to explain BAL-quasar outflows. In this model, the wind is launched from the disk at $\sim$\,10$^{16}$--10$^{17}$~cm \citep{murray1995accretion,proga2000dynamics} and BALs are generally seen when the system is observed at large inclination angles, so the line-of-sight of the observer intersects the wind (e.g. \citealt{risaliti10,ak2014dependence,leighly15}). The efficiency of line driving falls with the increasing ionization state of the wind. To address the issue of wind overionization, the model invokes ``shielding-gas'' to prevent the wind from becoming too ionized by the \hbox{X-ray} and extreme-UV (EUV) radiation from the innermost accretion disk/corona. 

In the shielding-gas model, AGNs with outflowing winds should be \hbox{X-ray} weak due to \hbox{X-ray} absorption by the shielding gas. Studies (e.g. \citealt[][]{green96,brandt2000nature,gallagher2002x,gallagher2006exploratory,fanll,gibson2009catalog}) have indeed demonstrated that BAL quasars are typically \hbox{X-ray} weak; in some cases \hbox{X-ray} spectroscopic studies directly show this \hbox{X-ray} weakness is due to \hbox{X-ray} absorption. Furthermore, the level of observed \hbox{X-ray} weakness is correlated with the absorption strength and maximum velocity of the UV \civ~BAL (e.g. \citealt{gibson2009catalog,wu2010x}), indicating that the \hbox{X-ray} absorption indeed affects the properties of the UV wind. \hbox{X-ray} absorption in BAL quasars is complex; i.e., not just a simple photoelectric absorption cutoff but an ionized or partially covering absorber. The estimated hydrogen column densities are generally in the range 10$^{22}$--5$\times$10$^{23}$~cm$^{-2}$ \citep[e.g.][]{gallagher2002x,gallagher2006exploratory}. Notable variability of \hbox{X-ray} absorption in a few BAL quasars has been detected on timescales of years (e.g. \citealt{gallagher2004dramatic,miller2006x,chartas2009confirmation,saez2012long}), suggestive of the dynamical nature of shielding gas which might be subject to rotational and inward/outward translational motions.

Another school of thought (e.g. \citealt{junkkarinen1983non,weymann1985broad,hamann2013extreme}) explains the observations based on clumping of the wind; such local density enhancements could significantly reduce the ionization parameter of the plasma without the need for shielding gas or large total column densities. \citet{baskin2014radiation} discuss the framework of a highly clumped wind considering the effects of radiation pressure confinement. 

Hitherto, key sample-based studies that have addressed the issue of BAL disappearance include the works of \citet{ak2012}, \citet{mcgraw2017broad}, and \citet{decicco2018}. \citet{ak2012} presented the first statistical analysis of \civ~BAL disappearance using 19 objects observed on 1.1--3.9~yr rest-frame timescales. These objects were selected from a sample of 582 BAL quasars that were drawn from a project making use of data from the Sloan Digital Sky Survey-I/II (henceforth SDSS; \citealt{york2000sloan}) and the SDSS-III Baryon Oscillation Spectroscopic Survey (BOSS; \citealt{eisenstein2011sdss,dawson2012baryon}) with at least two epochs of observation for each source, allowing comparison between them to investigate BAL variability. They identified ten cases out of the 19 BAL quasars which transformed from being BAL quasars in SDSS to being non-BAL quasars in the BOSS; these were the first reported examples of such transformations, hereafter referred to as ``transforming BAL quasars''. Moreover, they concluded that BAL troughs that disappear tend to have small-to-moderate EWs, comparatively shallow depths, and high outflow velocities. 

\citet{mcgraw2017broad} carried out an investigation of \civ~and \siv~BAL disappearance and emergence using a sample of 470 BAL quasars with rest-frame observing baselines ranging from 0.10--5.25~yr and at least three spectroscopic epochs for each source, from the SDSS, BOSS, and Time Domain Spectroscopic Survey (TDSS; \citealt{morganson2015time}), and report on the \civ~and \siv~variability of 14 disappearing BALs and 18 emerging BALs. In a recent study, which is the largest in terms of sample size, \citet{decicco2018} performed an analysis of 1319 BAL quasars and identified 30 transforming BAL quasars over a rest-frame time-scale of 0.28--4.9~yr, and determined the fraction of sources with disappearing BALs to be $5.1^{+0.7}_{-0.6}$ per cent. They estimate the average lifetime of a BAL along our line of sight is \hbox{$\approx$\,80--100~yr}, which is in agreement with the accretion-disk orbital time at distances of $\approx$\,0.1~pc. Other previous studies of BAL appearance/disappearance, often focusing on single objects or just a few objects, include the works of~\citet{junkv01}, \citet{maf2002}, \citet{lundgren2007broad}, \citet{hamann2008emergence}, \citet{leighly2009}, \citet{krongold2010transition}, \citet{hall2011implications}, \citet{capellupo2012variability}, \citet{vivek2012dynamically}, \citet{zhang2015}, \citet{saturni2016multi}, \citet{vivek2016}, and \citet{luweijian2018}.

In this paper, we analyze and interpret exploratory \hbox{\chandra}~\hbox{X-ray} observations of a sample of eight radio-quiet quasars that have undergone transformations from being BAL quasars in SDSS to being non-BAL quasars in the BOSS. In addition, we acquired contemporaneous optical/UV spectra of these eight sources with the \hbox{\gemini}~and ARC 3.5-m telescopes. A multi-wavelength perspective on these transforming BAL quasars can provide crucial clues about their nature.

The paper is organized as follows: in Section~\ref{sec:selection}, we present the details of sample selection, the target quasar properties, and details of their observations; in Section~\ref{sec:DATA_analysis} we describe the \hbox{\chandra}~\hbox{X-ray} data analysis, the optical/UV spectral reduction process, and the method used to probe \civ~BAL disappearance; in Section~\ref{sec:Results_Discussion} we present the results and discuss our findings. We conclude in Section~\ref{sec:summary} by summarizing our results and discussing future prospects. Throughout this work, we assume a cosmology with $H_{0}\,=\,70$~km~s$^{-1}$~Mpc$^{-1}$, $\Omega_{\rm M}\,=\,0.3$, and $\Omega_{\Lambda}\,=\,0.7.$ The EWs and times are in the quasar rest frame unless otherwise noted.

\section{SAMPLE SELECTION, SAMPLE PROPERTIES, AND OBSERVATIONS}
\label{sec:selection}
\subsection{Sample selection and properties}
\label{sec:sampleselection}

The targets of interest in the current study are the eight radio-quiet transforming BAL quasars presented by \citet{ak2012}; the other two transforming BAL quasars from the \citet{ak2012} sample were radio loud. They were chosen from a sample of 582 SDSS BAL quasars having rest-frame multi-year optical/UV spectroscopic coverage and represent some of the most exceptional BAL variations found within this large sample. These BAL quasars showed the disappearance of \civ~and all other troughs seen in SDSS-I/II when juxtaposed with the later and higher quality SDSS-III BOSS spectra.

\begin{figure}
\centerline{
\includegraphics[scale=0.35]{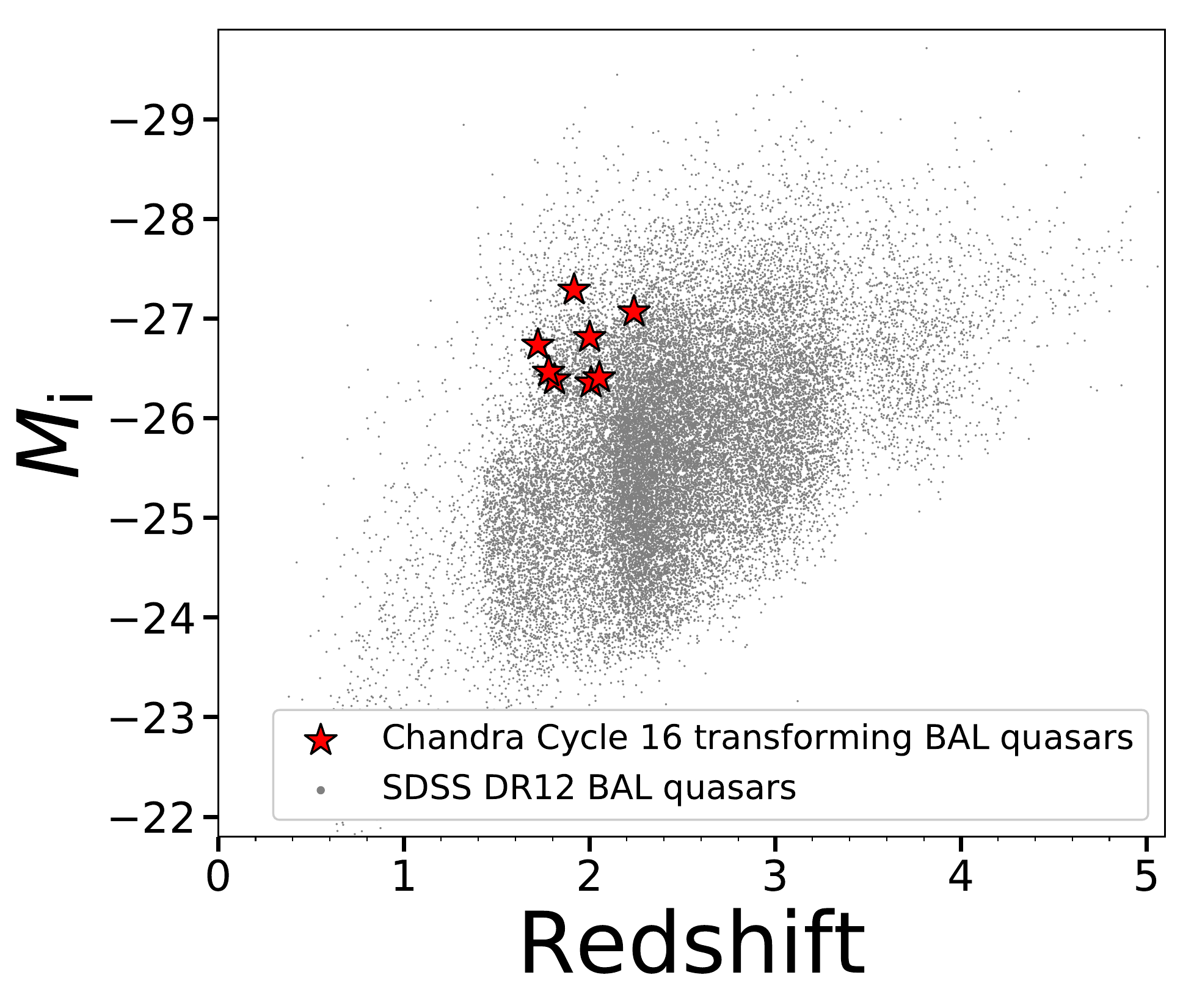}
}
\caption{
The stars indicate the redshifts and SDSS absolute
$i$-band magnitudes of the eight transforming BAL quasars in our study. For comparison, the gray dots represent the 29,580 BAL quasars from
the SDSS DR12 quasar catalog~\citep{paris2017}.}
\label{Figure1}
\end{figure}

The transforming BAL quasars in this study have redshifts of $z$\,=\,1.72--2.24 with an average redshift of 1.94. The redshifts and absolute $i$-band magnitudes of these targeted quasars are compared to those of the 29,580 BAL quasars from the SDSS DR12 quasar catalog~\citep{paris2017} in Fig.~\ref{Figure1}. Our targets belong to the category of HiBALs, as they have \civ~BALs but do not have any BALs from lower-ionization species present in their original SDSS-I/II observation spectra. Their \civ~BAL troughs generally show small-to-moderate EWs, relatively shallow depths, and high outflow velocities as indicated in Table~\ref{tab:table-1}. All of these quasars have spectra that provide complete coverage of the \civ~region.

Our targets have bright optical fluxes in the range \hbox{$m_{i}$\,=\,18.0--18.7} permitting suitably sensitive \hbox{\chandra}~observations to be obtained economically. The basic \hbox{X-ray}, optical, and radio properties for these sources are listed in \hbox{Table~\ref{tab:table-2}}.

\setlength{\tabcolsep}{0.20em}
\begin{table*}
\caption{\bf Optical/UV spectroscopic observations of transforming BAL quasars \label{tab:table-1}}
\textwidth=7.0in
\textheight=8.0in
\vspace*{0.30in}

\noindent  
\centering
\resizebox{\textwidth}{!}{\begin{tabular}{ccccccccccc} \hline \nonumber
Object Name &  Facility & Plate-MJD-Fiber & Observation & BI & EW(1$\sigma$) & Velocities & Max Depth & Difference in days & Status\\  
(J2000) & & &  Date & (km~s$^{-1}$) & (\angstrom) & (Max,Min) & & \hbox{X-ray} \& Optical& \\
&&&&&& (km s$^{-1}$) & & (Rest frame)&\\
(1)&(2)&(3)&(4)&(5)&(6)&(7)&(8)&(9)&(10)&\\\hline

 074650.59$+$182028.7  & SDSS & 1582-52939-95 & 2003-10-27 & 1260 & 9.05(0.21) & ($-24875,-17526$)  & 0.33 & 1514.04 & First observation\\	
	 & BOSS & 4492-55565-828 & 2011-01-04 &  0 & $<4.25$ & - & - & 614.73 & Disappeared\\
	 & \hbox{\gemini} &  & 2016-02-08 & 0 & $<5.76$  & - & - & $-23.29$ & Disappeared\\
\\
 085904.59$+$042647.8  & SDSS & 1192-52649-291 & 2003-01-10 & 781 & 4.82(0.50) & ($-18858,-16037$)  & 0.48 & 1681.90 & First observation\\\
	 & BOSS & 3817-55277-538 & 2010-03-22 & 0 & $<1.82$ & - & - & 752.67 & Disappeared\\
	 & BOSS & 3814-55535-928 & 2010-12-05 & 0 & $<1.76$ & - & - & 660.50 & Disappeared\\
	 & \hbox{\gemini} & & 2016-02-11 & 0 & $<0.31$ & - & - & $-13.88$ & Disappeared\\
\\
093620.52$+$004649.2  & SDSS & 476-52027-442 & 2001-04-28 & 0  & $<2.47$ & - & - & 1864.33 & First observation\\
	 & SDSS & 476-52314-444 & 2002-02-09 & 925 & 5.93(0.33) & ($-17212,-13494$) & 0.44  &1758.82 & BAL appearance\\
	 & BOSS & 3826-55563-542 & 2011-01-02 & 0 &  $<0.74$ & - & - & 564.33 & Disappeared\\
	 & \hbox{\gemini} & & 2015-11-10 & 0 & $<1.59$ & - & - & 98.90 & Disappeared\\
\\
114546.22$+$032251.9  & SDSS & 514-51994-458 & 2001-04-28 & 338 & 3.00(0.32) & ($-12343,-9448$)  & 0.25 & 1805.31 & First observation\\
	 & BOSS & 4766-55677-50 & 2011-04-26 & 0 & - & - & - & 581.73 & Disappeared\\
	 & \hbox{\gemini} & & 2016-03-15 & 0 & - & - & - & $-11.96$  & Disappeared\\
\\
133211.21$+$392825.9  & SDSS & 2005-53472-330 & 2005-04-12 & 725 & 4.92(0.37) & ($-20999,-17631$) & 0.40 & 1241.64 & First observation\\
	 & BOSS & 4708-55704-412 & 2011-05-23 & 0 & $<0.78$ & - & - & 509.84 & Disappeared\\
	 & ARC 3.5-m & & 2015-06-18 & 0 & $<1.36$ & - & - &  21.97 & Disappeared\\
\\
142132.01$+$375230.3   & SDSS & 1380-53084-13 & 2004-03-20 & 846 & 5.04(0.34) & ($-16612,-13996$) & 0.56 & 1517.63  & First observation\\
	 & BOSS & 4712-55738-30 & 2011-06-26 & 0 & $<1.33$ & - & - & 562.95  & Disappeared\\
	 & BOSS & 4713-56044-532 & 2012-04-27 & 0 & $<2.13$ & - & - & 456.47 & Disappeared\\
	 & \hbox{\gemini} & & 2016-03-03 & 0 & $<0.55$ & - & - & $-56.60$ & Disappeared \\
\\
152149.78$+$010236.4  & SDSS & 313-51673-339 & 2000-05-09 & 825 & 6.07(0.45) & ($-23383,-18231$) & 0.34 & 1701.54 & First observation  \\
	 & SDSS & 2953-54560-209 & 2008-04-04 & 0 & $<1.18$ & - & - & 810.49 & Disappeared \\
	 & BOSS & 4011-55635-166 & 2011-03-15 & 0 & $<0.60$ & - & - & 478.70 & Disappeared \\
	 & \hbox{\gemini} & & 2015-04-15 & 0 & $<0.02$ & - & - & 18.21 & Disappeared \\
\\
152243.98$+$032719.8  & SDSS & 592-52025-254 & 2001-04-26 & 309 & 2.44(0.20) & ($-14234,-12099$) & 0.28 & 1700 & First observation \\
	 & BOSS & 4803-55734-442 & 2011-06-22 & 0 & - & - & - & 463.67 & Disappeared \\
	 & \hbox{\gemini} & & 2015-04-15 & 0 & - & - & - & $-1.33$ & Disappeared  \\\hline
\end{tabular}} \\
\begin{flushleft}
      \small
Cols.: (1) Object name, (2) Observing facility, (3) Plate-MJD-Fiber, (4) Date of Observation, (5) Balnicity Index of the \civ~line, (6) Equivalent width of the \civ~line and 1$\sigma$ error. We provide the 90 per cent confidence upper limits on the EW in cases where the BAL trough had disappeared. In the cases of SDSS J074650.59$+$182028.7 and SDSS J085904.59$+$042647.8, NALs are blended with the disappearing BAL trough resulting in the over estimation of the EW upper limits. In the cases of SDSS J114546.22$+$032251.9 and SDSS J152243.98$+$032719.8, the bluer portion of the \civ~emission in the BOSS and \gemini~spectra extends into the BAL--trough wavelength limits resulting in negative EW; for these cases we do not quote the upper limit on the EW, (7) Maximum and minimum velocities of the \civ~line, (8) The maximum fractional depth of the BAL trough below the continuum averaged over five pixels. In the case of SDSS J074650.59$+$182028.7, we excluded the NAL doublets blended with the disappearing \civ~BAL trough in the determination of the Max depth, (9) Gap in rest-frame days between \chandra~and corresponding optical observation; a negative sign indicates that the optical observation has been carried out after the \chandra~observation, (10) The status of the BAL trough; First observation - denotes the first SDSS observation, Disappeared - denotes that the BAL trough is absent in BOSS or \gemini~or ARC 3.5-m observation. In the case of SDSS J093620.52$+$004649.2, the BAL appeared in the second epoch of SDSS observation.
      
    \end{flushleft}
\end{table*}

\setlength{\tabcolsep}{0.35em}
\begin{table*}
\centering
\caption{\bf \hbox{X-ray}, optical, and radio properties \label{tab:table-2}}
\textwidth=5.0in
\textheight=6.0in
\vspace*{0.12in}
\begin{adjustbox}{center}
\noindent  

\begin{tabular}{ccccccccccccccc} \hline \nonumber
Object Name (SDSS J) & $z$ & $m_\text{i}$ & $M_\text{i}$ & $N_\text{H}$ & Count & $F_\text{0.5--2 keV}$ & $f_\text{2 keV}$ & log$L_\text{X}$ & $f_\text{2500~\AA}$ & log$L_{\nu}$ & $\alpha_{ox}$ & $\Delta\alpha_{ox}(\sigma)$ & $R$  \\
	&	&	&&&	Rate&	& &   (2--10 keV) & & (2500~\AA) & && \\
$(1)$&$(2)$&$(3)$&$(4)$&$(5)$&$(6)$&$(7)$&$(8)$&$(9)$&$(10)$&$(11)$&$(12)$&$(13)$&$(14)$\\\hline

074650.59$+$182028.7 & 1.92 & 17.96 & $-27.36$ & 4.17 & $1.63^{+1.27}_{-0.77}$ & $0.94^{+0.79}_{-0.53}$ & $4.11^{+3.43}_{-2.30}$  & 44.46 & 2.24 $\pm$ 0.22 & 31.32 $\pm$ 0.10 & $-1.82^{+0.10}_{-0.08}$ & $-0.14$(0.89)&\textless 2.17 \\
085904.59$+$042647.8 & 1.81 & 18.74 & $-26.45$ & 4.11 & $\textless 1.0$ & \textless 0.57 &\textless 2.39  &\textless 44.18 & 0.75 $\pm$ 0.07 & 30.80 $\pm$ 0.10 & \textless$-1.72$ & \textless$-0.12$(0.64)&  \textless 4.46\\

093620.52$+$004649.2 & 1.72 & 18.26 & $-26.81$ & 3.85 & $4.38^{+1.72}_{-1.28}$ & $2.51^{+1.24}_{-1.05}$ & $10.21^{+5.05}_{-4.27}$  & 44.77 & 1.15 $\pm$ 0.12 & 30.95 $\pm$ 0.10 & $-1.56^{+0.07}_{-0.06}$ & $0.07$(0.41)& \textless 2.94\\
114546.22$+$032251.9 & 2.01 & 19.01 & $-26.42$ & 2.19 & $2.54^{+0.95}_{-0.71}$ & $1.40^{+0.67}_{-0.58}$ & $6.31^{+3.02}_{-2.58}$  & 44.68 & 0.72 $\pm$ 0.07 & 30.87 $\pm$ 0.10 & $-1.56^{+0.07}_{-0.06}$ & 0.06(0.33)&\textless 5.74\\

133211.21$+$392825.9 & 2.05 & 19.01 & $-26.46$ & 0.89 & $4.31^{+1.17}_{-0.94}$ & $2.31^{+0.93}_{-0.86}$ & $10.49^{+4.24}_{-3.89}$  & 44.91 & 0.75 $\pm$ 0.08 & 30.89 $\pm$ 0.10 & $-1.48^{+0.06}_{-0.06}$ & 0.14(0.79)&\textless 5.21\\
142132.01$+$375230.3 & 1.78 & 18.61 & $-26.54$ & 0.92 & $1.79^{+1.05}_{-0.70}$ & $0.96^{+0.63}_{-0.47}$ & $3.99^{+2.63}_{-1.96}$  & 44.38 & 1.02 $\pm$ 0.10 & 30.92 $\pm$ 0.10 & $-1.69^{+0.09}_{-0.07}$ & $-0.07$(0.38)&\textless 3.66\\
152149.78$+$010236.4 & 2.24 & 18.45 & $-27.22$ & 4.24 & $1.39^{+0.92}_{-0.59}$ & $0.81^{+0.59}_{-0.42}$ & $3.89^{+2.83}_{-2.02}$  & 44.55 & 2.04 $\pm$ 0.21 & 31.40 $\pm$ 0.10 & $-1.81^{+0.09}_{-0.07}$ & $-0.12$(0.79)&\textless 3.96\\

152243.98$+$032719.8 & 2.00 & 18.56 & $-26.86$ & 3.81 & $1.39^{+0.92}_{-0.59}$ & $0.80^{+0.58}_{-0.41}$ & $3.57^{+2.60}_{-1.86}$  & 44.43 & 1.31 $\pm$ 0.13 & 31.12 $\pm$ 0.10 & $-1.75^{+0.09}_{-0.07}$ & $-0.10$(0.59)&\textless 3.88\\\hline 

     \end{tabular} 
\end{adjustbox}

\begin{flushleft}
\small
\noindent Column (1): SDSS J2000 equatorial coordinates.
\noindent Column (2): Redshift of the quasar~\citep{Hewett}.
\noindent Column (3): Apparent extinction-corrected $i$-band magnitude, $m_{i}$, from the 
SDSS DR12 quasar catalog~\citep{paris2017}.
\noindent Column (4): Absolute $i$-band magnitude, $M_{i}$, from the 
SDSS DR12 quasar catalog~\citep{paris2017}.
\noindent Column (5): Galactic neutral hydrogen column density obtained 
with the {\it Chandra} COLDEN tool, in units of $10^{20}$~cm$^{-2}$.
\noindent Column (6): Count rate in the observed-frame soft \hbox{X-ray} band ($0.5-2.0$~keV), in units of $10^{-3}$~s$^{-1}$, obtained from our \chandra~observations. 
\noindent Column (7): Galactic absorption-corrected flux in the observed-frame soft \hbox{\hbox{X-ray}} band, in units of $10^{-14}$~erg~cm$^{-2}$~s$^{-1}$, obtained from our \chandra~observations. The error bars include i) uncertainty in flux propagated from the count rate and ii) uncertainty due to variability ($\approx$\,30 per cent) based on an estimate from \citet{saez2012long}, which are added in quadrature. 
\noindent Column (8): Flux density at rest-frame 2~keV, in units of $10^{-32}$ erg~cm$^{-2}$~s$^{-1}$~Hz$^{-1}$, obtained from our \chandra~observations. 
\noindent Column (9): Logarithm of the Galactic absorption-corrected quasar luminosity in the rest frame 2--10~keV band, obtained from our \chandra~observations. 
\noindent Columns (10) and (11): Flux density at rest-frame 2500~\AA~in units of $10^{-27}$~erg~cm$^{-2}$~s$^{-1}$~Hz$^{-1}$ and the logarithm of the monochromatic luminosity at rest-frame 2500~\AA, respectively, obtained from \citet{shen2011}. The error bars include i) uncertainty in flux obtained from the pixel errors using a window of 50 pixels centered on (1+$z$)~2500~\AA~and ii) uncertainty due to variability ($\approx$\,10 per cent) based on an estimate from \citet{macleod2010modeling}, which are added in quadrature. 
\noindent Column (12): The \hbox{X-ray}-to-optical power-law slope as defined in Footnote~2. The errors from $f_\text{2 keV}$ and $f_\text{2500~\AA}$ are propagated using the numerical method described in section 1.7.3 of \citet{lyons1991practical}.
\noindent Column (13): $\Delta\alpha_{\rm ox}$ as defined in Footnote~2. The statistical significance of the \daox~difference from Footnote~2, given in parentheses, is in units of $\sigma$, where $\sigma$ is 
given in Table~5 of \citet{steffen2006x} as the RMS $\alpha_{\rm ox}$ for 
various luminosity ranges. Here, $\sigma$\,=\,0.198 for 
30\,$<$\,log$L_{2500\mbox{\rm~\scriptsize\AA}}$\,$<$\,31 and $\sigma$\,=\,0.146 for 
31\,$<$\,log$L_{2500\mbox{\rm~\scriptsize\AA}}$\,$<$\,32.
\noindent Column (14): The radio-loudness parameter is given by 
$f_{5\;{\rm GHz}}$/$f_{4400\mbox{\rm~\scriptsize\AA}}$
where the denominator, $f_{4400\mbox{\rm~\scriptsize\AA}}$, was found via extrapolation from 
$f_{2500\mbox{\rm~\scriptsize\AA}}$ using an optical/UV power-law slope of $\alpha_\nu$\,=\,$-0.5$~\citep{richstone1980spectral}. 
The numerator, $f_{5\;{\rm GHz}}$, was found using a radio power-law slope of $\alpha_\nu$\,=\,$-0.8$ and a flux limit at 20 cm, $f_{20\;{\rm cm}}$, of $0.25$\,+\,3$\sigma_{rms}$, where $\sigma_{rms}$ is the rms noise of the FIRST survey at the source position and 0.25~mJy is to account for the CLEAN bias \citep{white}. All the upper limits are well below the commonly used criterion for radio-quiet objects ($R\,<\,10$). 
\end{flushleft}
\end{table*}

\setlength{\tabcolsep}{0.40em}
\begin{table*}
\caption{\bf \hbox{\chandra}~observation log and \hbox{X-ray} counts\label{tab:table-3}}
\textwidth=5.0in
\textheight=6.0in
\vspace*{0.20in}
\noindent  
\centering
\begin{tabular}{ccccccccccc} \hline \nonumber
Object Name (SDSS J) & ObsID  & Date & Exposure Time & Full-Band Cts. & Soft-Band Cts. & Hard-Band Cts. & Band & $\Gamma_{\text{eff}}$\\
& & & (ks) & (0.5--8.0~keV) & (0.5--2.0~keV) & (2.0--8.0~keV) & Ratio \\
$(1)$&$(2)$&$(3)$&$(4)$&$(5)$&$(6)$&$(7)$&$(8)$&$(9)$ 
&\\\hline  
 074650.59$+$182028.7 & 17021  & 2015 Dec 03 & $2.54$ & $9.51^{+4.21}_{-3.04}$ & $4.15^{+3.22}_{-1.95}$ &$5.46^{+3.50}_{-2.27}$ & $1.32^{+1.02}_{-0.69}$ & $0.67^{+0.72}_{-0.54}$\\	
 085904.59$+$042647.8 & 17023  & 2016 Jan 04 & $3.92$ & $3.09^{+2.99}_{-1.69}$ & $\textless 4.07$ &$2.13^{+2.72}_{-1.36}$ & $\textgreater 0.52$ &$\textless 1.56$ \\
 093620.52$+$004649.2 & 17022  & 2015 Mar 16 & $2.61$ & $13.76^{+4.82}_{-3.67}$ & $11.44^{+4.50}_{-3.33}$ 
 &$2.15^{+2.72}_{-1.36}$& $0.19^{+0.24}_{-0.13}$       & $2.55^{+1.06}_{-0.80}$\\
 114546.22$+$032251.9 & 17029  & 2016 Feb 09 & $4.91$ & $16.88^{+5.22}_{-4.08}$& $12.46^{+4.64}_{-3.49}$ 
 &$4.30^{+3.27}_{-2.01}$ & $0.34^{+0.28}_{-0.18}$       & $1.93^{+0.70}_{-0.57}$\\
 133211.21$+$392825.9 & 17028  & 2015 Aug 24 & $4.83$ & $25.48^{+6.13}_{-5.02}$& $20.84^{+5.65}_{-4.53}$ 
 &$4.35^{+3.27}_{-2.01}$ & $0.21^{+0.16}_{-0.10}$       & $2.38^{+0.64}_{-0.55}$\\
 142132.01$+$375230.3 & 17027  & 2015 Oct 07 & $3.49$ & $7.38^{+3.86}_{-2.66}$ & $6.24^{+3.65}_{-2.43}$ 
 &$\textless 4.26$ & $\textless 0.68$      & $\textgreater 1.25$\\
 152149.78$+$010236.4 & 17026  & 2015 Jun 12 & $3.73$ & $6.29^{+3.67}_{-2.46}$ & $5.18^{+3.44}_{-2.20}$ 
& $\textless 4.23$ & $\textless 0.82$        & $\textgreater 1.13$ \\
 152243.98$+$032719.8 & 17024  & 2015 Apr 12 & $3.73$ & $11.62^{+4.53}_{-3.37}$ & $5.18^{+3.44}_{-2.20}$ 
& $6.56^{+3.71}_{-2.50}$ & $1.27^{+0.88}_{-0.61}$        & $0.70^{+0.64}_{-0.50}$\\
\hline 
     \end{tabular} \\
\begin{flushleft}
\small
\item {
\noindent Column (1): The SDSS J2000 equatorial coordinates for the quasar.
\noindent Column (2): \hbox{\chandra}~observation ID.
\noindent Column (3): Date of observation of the target with \hbox{\chandra}.
\noindent Column (4): Background-flare cleaned effective exposure time.
\noindent Columns (5), (6), (7): The \hbox{X-ray} data analysis was carried out using the procedure in \citet{Luo2015}. Errors on the \hbox{X-ray} counts were calculated using Poisson statistics corresponding to the 1$\sigma$ significance level using \citet{Gehrels1986}. A source is considered undetected if the $P_{\rm B}$ value, defined via Equation (3) in Section~\ref{sec:chandraanalysis}, in a band is $>$\,0.01, in which case an upper limit on the source counts was derived. \noindent Column (8): Band ratio is the hard band to soft band ratio. \noindent Column (9): The effective power-law photon index, $\Gamma_{\text{eff}}$.}
\end{flushleft}
\end{table*}

\subsection{\textbf{\hbox{\chandra}}~observations}
\label{sec:chandra_obs}

The \hbox{\chandra}~observations of our eight transforming BAL quasars were performed between 2015 March 16 and 2016 February 09, using the Advanced CCD Imaging Spectrometer (ACIS; \citealt{garmire2003advanced}) spectroscopic array (ACIS-S). Our targets were placed on the back-illuminated S3 detector for high-resolution imaging, following standard practice. VFAINT mode was used to minimize the particle background and help distinguish between true \hbox{X-ray} events and events that are most likely associated with cosmic rays. The exposure times for our targets were \hbox{2.54--4.91~ks} and were set to obtain highly significant detections provided our targets have \aox~$\geq$\,$-1.9$, which corresponds to the targets being up to $\approx$\,5 times weaker in \hbox{X-rays} than expectations for a typical (non-BAL) radio-quiet quasar of similar optical/UV luminosity.\footnote{\label{footnote:one} The \hbox{X-ray}-to-optical power-law slope, given by
\begin{equation}
    \alpha_{\rm ox} = \frac{{\rm log}(f_{\rm 2\;keV} / 
    f_{2500\mbox{\rm~\scriptsize\AA}})}{{\rm log}(\nu_{\rm 2\;keV} / \nu_{2500\mbox{\rm~\scriptsize\AA}})}
\end{equation}
\noindent Also, $\Delta\alpha_{\rm ox}$ is defined as
\begin{equation}
    \Delta\alpha_{\rm ox} = \alpha_{\rm ox(measured)} - \alpha_{\rm ox(expected)}
\end{equation}
The expected value of $\alpha_{\rm ox}$ for a typical radio-quiet quasar is calculated from the established $\alpha_{\rm ox}$--$L_{2500\mbox{\rm~\scriptsize\AA}}$ correlation given as 
Equation (3) of J07.} In these calculations, the standard \hbox{$\alpha_{\rm ox}$--$L_{2500\mbox{\rm~\scriptsize\AA}}$} correlation (e.g. \citealt{just2007x}, hereafter J07) was adopted. These calculations also accounted for the time-dependent degradation of the ACIS-S low-energy quantum efficiency. A summary of the \hbox{\chandra}~observations of our targets is presented in Table~\ref{tab:table-3}. 

We have furthermore searched for sensitive archival \hbox{X-ray} observations of our targets performed prior to the \chandra~observations. No such observations exist.

\subsection{Optical/UV observations}
\label{sec:optical_obs}

Prior to this study, the transforming BAL quasars had optical/UV spectral observations from SDSS-I/II and the SDSS-III BOSS. We obtained additional data from the \hbox{\gemini}~and ARC 3.5-m telescopes for these sources in order to set contemporaneous constraints upon rest-frame UV BALs and continuum emission to support the interpretation of our \hbox{\chandra}~data. These supporting optical/UV spectral observations were within 1--90 rest-frame days of our \hbox{\chandra}~observations with a median separation of 20 rest-frame days.  

The SDSS observations were carried out using a \hbox{2.5-m} Ritchey-Chretien telescope located at the Apache Point Observatory in New Mexico~\citep{gunn20062}. The fully reduced spectra were obtained from the publicly available SDSS Data Release (DR12). The SDSS spectra, in general, cover a broad wavelength range spanning from \hbox{3800--9200~angstrom}~at a spectral resolution of $R$\,$\approx$\,2000. The BOSS observations used the same telescope as SDSS, but utilized a different spectrograph with extended wavelength coverage spanning between \hbox{3650--10400~\angstrom}~and an identical spectral resolution to SDSS~\citep{smee2013multi}. The ARC 3.5-m observation of SDSS J133211.21$+$392825.9 was obtained using the Dual Imaging Spectrograph. We obtained three exposures of 600~s, using the B400/R300 grating (R\,$\sim$\,800 and wavelength coverage of \hbox{3400--9200}~\angstrom) with a 1.5$\arcsec$ slit. The seeing was approximately 1.58$\arcsec$ on this night, and the exposures were obtained at a mean airmass of 1.198. Spectra of the spectrophotometric standard star Feige 66 were also obtained for flux-calibration and removal of atmospheric absorption, and \hbox{HeNeAr} lamps were used for wavelength calibration. These spectra were bias and flat-field corrected, wavelength and flux-calibrated, and corrected for atmospheric extinction using standard IRAF\footnote{IRAF is distributed by the National Optical Astronomy Observatory, which is operated by the Association of Universities for Research in Astronomy (AURA) under a cooperative agreement with the National Science Foundation.} procedures. The~\hbox{\gemini}~observations were obtained using the \hbox{\gemini}~Multi-Object Spectrograph (GMOS) on the 8.1-m Gemini North Cassegrain telescope situated on the summit of Mauna Kea, Hawai'i~\citep{hook2004gemini}. The data were collected during the Gemini 15A, 15B, and 16A observing cycles. The spectral wavelength coverage for the \hbox{\gemini}~data spans a narrower range than SDSS, extending from \hbox{3200--6000}~\angstrom. The exposure times for our \hbox{\gemini}~observations range between \hbox{1168--2408}~s with a mean exposure time of $\approx$\,1626~s. The \gemini~data were reduced using the IRAF \hbox{\gemini}~GMOS package and standard reduction methods. Each object was flux calibrated using spectrophotometric standard stars from the \hbox{\gemini}~archive, most of which were not taken on the same night as our spectra. To compare the \hbox{\gemini}~spectra to BOSS and SDSS, Equation 3 of~\citet{mor91} was used to convert the observed \hbox{\gemini}~air wavelengths (in~\angstrom) and flux densities to the vacuum wavelength system of SDSS and BOSS. Table~\ref{tab:table-1} provides a summary of the optical observations of these eight sources.

\section{DATA ANALYSIS}
\label{sec:DATA_analysis}
\subsection{\hbox{\chandra}~\hbox{X-ray} data analysis}
\label{sec:chandraanalysis}
The \hbox{X-ray} analysis of our \hbox{\chandra}~Cycle 16 targets was performed using the \hbox{\chandra}~Interactive Analysis of Observations (CIAO version 4.10) tools and CALDB version 4.7.9.\footnote{See http://cxc.harvard.edu/ciao/ for details on CIAO.} Each source was reprocessed using the \hbox{{\sc chandra\_repro}} script to apply the latest calibration. Background flares were removed using the {\sc deflare}
script with an iterative 3$\sigma$ clipping algorithm.
The final cleaned exposure times are listed in
Table~\ref{tab:table-3}. Images in the 0.5--8~keV (full),
0.5--2~keV (soft), and 2--8~keV (hard) bands were created for each source using the standard
\asca\ grade set (\asca\ grades 0, 2, 3, 4, and 6).
The images were searched for \hbox{X-ray} sources using {\sc wavdetect} \citep{freeman2002}
with a ``$\sqrt{2}$~sequence'' of wavelet scales (i.e.,\ 1, 1.414, 2,
2.828, and 4 pixels) and a false-positive probability threshold of 10$^{-6}$. If a source is detected in at least one band, then the {\sc wavdetect} position closest to the source's SDSS position is chosen to be the \hbox{X-ray} position.

The \hbox{X-ray}-to-optical positional offsets for the targets are small, and all of the eight sources are detected within 1\arcsec~of their SDSS positions. Aperture photometry was used to extract source counts in each of the three energy bands. A circular aperture with 2\arcsec-radius centered on the \hbox{\hbox{X-ray}} position and corresponding to encircled-energy fractions (EEFs) of 0.939, 0.959, and 0.907 in the full, soft, and hard bands, respectively, was used to extract the source counts. Background counts were extracted from within an annular region of inner radius 10\arcsec\ and outer radius 40\arcsec.

In order to assess the significance of the source signal \citep[e.g.][]{broos2007,xue2011,luo2013weak}, a binomial no-source probability, $P_{\rm B}$, defined as 
\begin{equation}
P_{\rm B}(X\ge S)=\sum_{X=S}^{N}\frac{N!}{X!(N-X)!}p^X(1-p)^{N-X}~
\end{equation}
was computed. Here, $S$ denotes the total number of counts in the
source-extraction region; $N=S+B$, where $B$
denotes the total number of counts in the background-extraction region; $p=1/(1+BACKSCAL)$,
where $BACKSCAL$ is {the ratio of the area of the background region to
that of the source region.} $P_{\rm B}$ is the probability of detecting the source counts by chance, assuming that there exists no real source at the location of interest.

A source is considered detected if the $P_{\rm B}$ value in a band is $\leq$\,0.01, corresponding to a \hbox{$\ge\,2.6\,\sigma$} detection. We note that the chosen threshold for $P_{\rm B}$ is warranted in our case as the position of interest is precisely specified in advance; therefore, only a single position needs to be considered in the source detection. Owing to the unmatched \hbox{X-ray} imaging capability of \chandra, the background in a source cell is very small. For example, the background is $\approx$\,0.1 counts for the 3-count source, SDSS J085904.59$+$042647.8. Given the low background of \chandra, a 3-count source at a pre-specified position is a highly significant detection for short exposures of $\approx$\,4~ks~(e.g. \citealt{shemmer06,vignali2006,kelly07,Luo2015}; see Equation 3). The 1$\sigma$ errors on the net counts were derived from the 1$\sigma$ errors on the extracted source and background counts \citep{Gehrels1986} using the prescription in Section 1.7.3 of \citet{lyons1991practical}. A source is considered undetected when the $P_{\rm B}$ value is larger than 0.01; in such cases an upper limit on the source counts was derived using the Bayesian approach of \citet{kraftrp} at a 90 per cent confidence level. 

Referring to Table~\ref{tab:table-2}, we obtained the Galactic absorption-corrected flux in the 
observed-frame soft \hbox{X-ray} band, $F_\text{0.5--2~keV}$, the flux density at rest-frame 2~keV, $f_\text{2~keV}$, and the logarithm of the Galactic absorption-corrected quasar luminosity in the rest-frame $2-10$~keV band using the \hbox{\chandra}~Portable, Interactive, Multi-Mission Simulator (PIMMS)\footnote{http://cxc.harvard.edu/toolkit/pimms.jsp.} tool (version 4.8c) with the \hbox{\chandra}~Cycle 16 instrument response, which accounts for the decrease in instrument efficiency with time; this calculation assumes an absorbed power-law model with a photon index $\Gamma$\,=\,2, which is typical for quasars and consistent with our data (see Section~\ref{sec:jointspectralanalysis}). The errors on $F_\text{0.5--2~keV}$ arise from i) flux uncertainty, which is propagated based on the error on the counts and ii) variability of the BAL quasar, for which we adopt a value of the \hbox{X-ray} variability amplitude of $\approx$\,30 per cent determined from \citet{saez2012long}. We find that the flux uncertainty is the dominant of the two error sources, and the total uncertainty is obtained by adding these two errors in quadrature. The $f_\text{2~keV}$ error is obtained by propagation of the error on $F_\text{0.5--2~keV}$. The flux density at rest-frame 2500~\AA~is obtained from \citet{shen2011}. The errors on $f_\text{2500~\AA}$ include i) flux uncertainty, obtained from the pixel errors using a window of 50 pixels centered on $(1+z)~\text{2500~\AA}$ and ii) variability of the BAL quasar, for which we adopt a value of the optical/UV variability amplitude of $\approx$\,10 per cent determined from \citet{macleod2010modeling}. We find that the error due to variability of the quasar is the dominant of the two error sources, and the total uncertainty is obtained by adding these two errors in quadrature. We have also derived the band ratio, defined as the hard band to soft band ratio, and the errors on the band ratio, which were calculated using Equation (1.31) in Section 1.7.3 of \citet{lyons1991practical}, corresponding to 1$\sigma$ significance. Subsequently, we have derived a \hbox{0.5--8~keV} effective power-law photon index ($\Gamma_{\rm eff}$) for each source using the band ratio between the hard-band and soft-band counts. The effective power-law photon index is measured using PIMMS assuming a Galactic-absorbed \citep{Dickey1990} power-law spectral model for each source. Table~\ref{tab:table-3} lists the source counts, band ratios, and $\Gamma_{\rm eff}$ values for our quasars.

\smallskip

\subsection{Optical/UV data analysis}
\label{sec:opticalanalysis}
\subsubsection{Continuum fitting and normalization}
\label{sec:continuumfit}

A number of operations were performed on the 28 spectra (listed in Table~\ref{tab:table-1}) prior to conducting continuum fits. The SDSS data-processing algorithm flags pixels that have significant residuals near prominent night-sky lines. The ``BRIGHTSKY'' mask column was examined, and such flagged pixels were removed from each spectrum (this step is not applicable to the spectra obtained from the \hbox{\gemini}~and ARC 3.5-m telescopes). The spectra were then corrected for Galactic extinction using a Milky Way extinction model \citep{Cardelli1989} for $R_{\mathrm{V}}$\,=\,3.1 using $A_{\mathrm{V}}$ values from \citet{schlafly2011measuring}. Observed wavelengths were translated to the rest-frame using redshifts from 
\citet{Hewett}.

\begin{figure*}
\begin{center}
\includegraphics[scale=0.440]{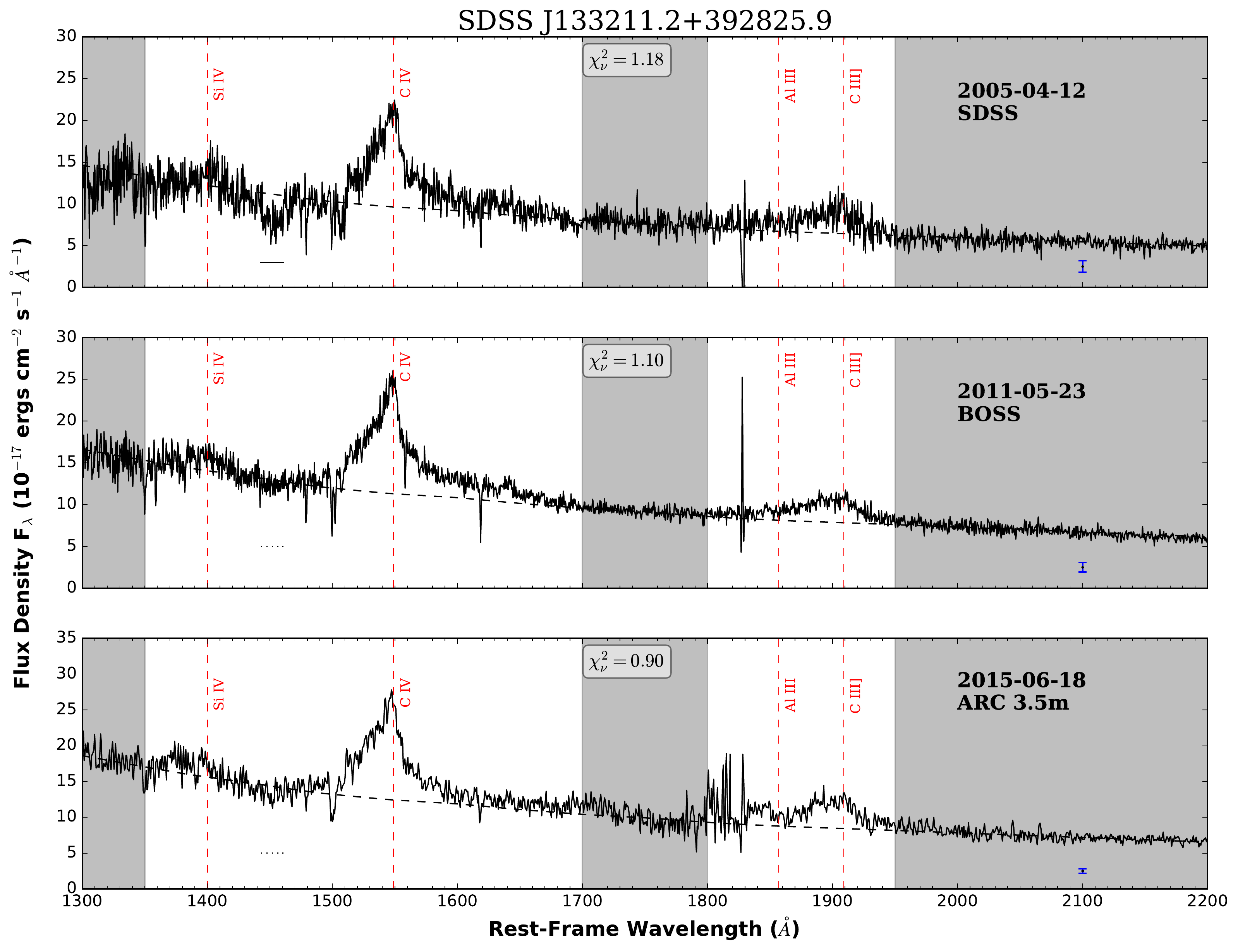}
\includegraphics[scale=0.440]{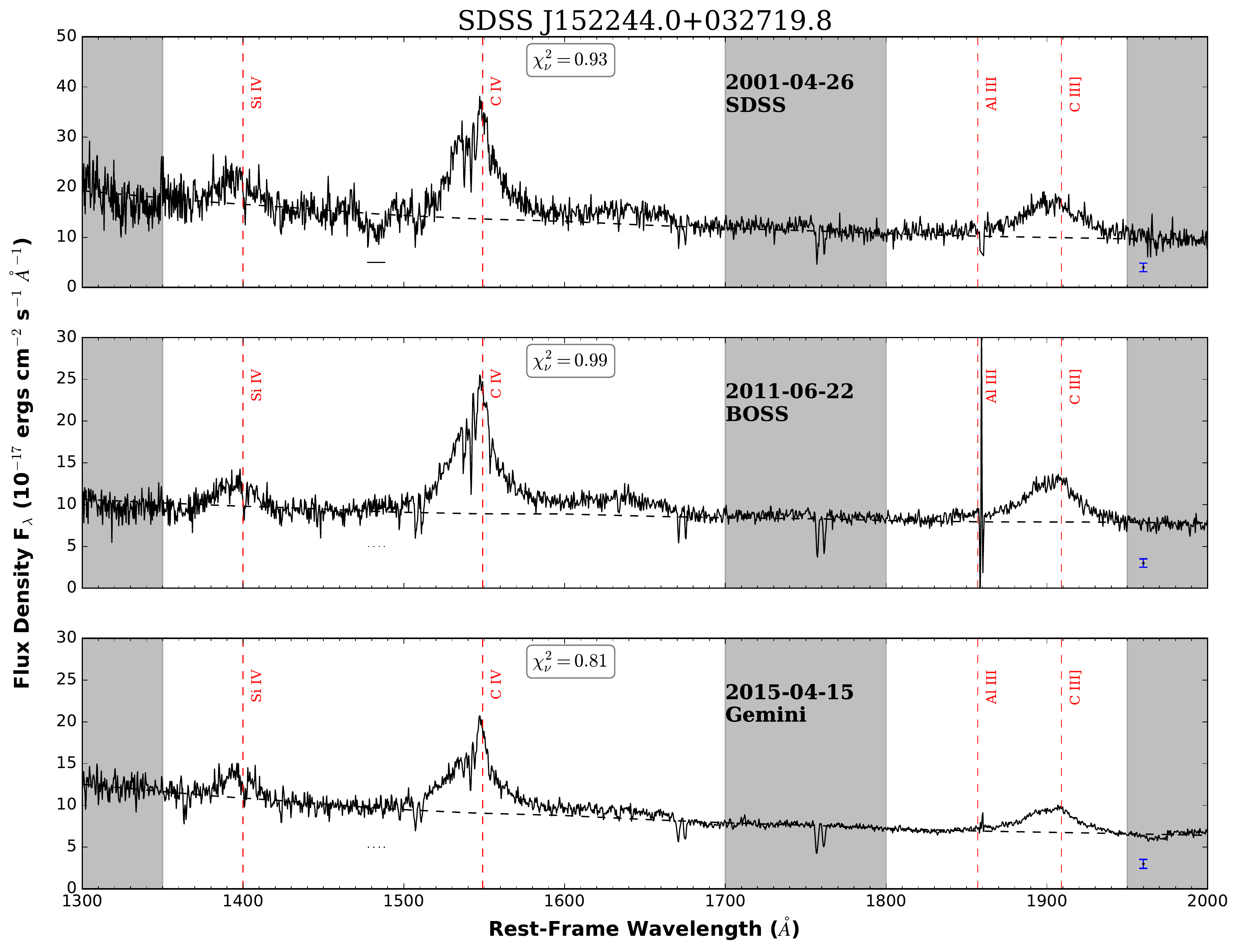}
\caption{
The continuum fits are shown as dashed lines for two representative cases (having an ARC 3.5-m observation and a \gemini~observation) of transforming BAL quasars. The reduced chi-squared statistic, $\chi^{2}_\nu$, is provided to indicate the fit quality in the RLF regions. The gray shaded areas indicate the RLF regions used to obtain the continuum fit. The horizontal black solid line extends over the range of the \civ~BAL region in the first SDSS spectrum; the BAL feature is not seen in the subsequent BOSS and \gemini~spectra as indicated by the corresponding black horizontal dotted lines. The median error bar on the flux density is shown at the bottom-right of each spectrum.
\label{Figure2}
}
\end{center}

\end{figure*}

\begin{figure*}
\begin{center}
\includegraphics[scale=0.75]{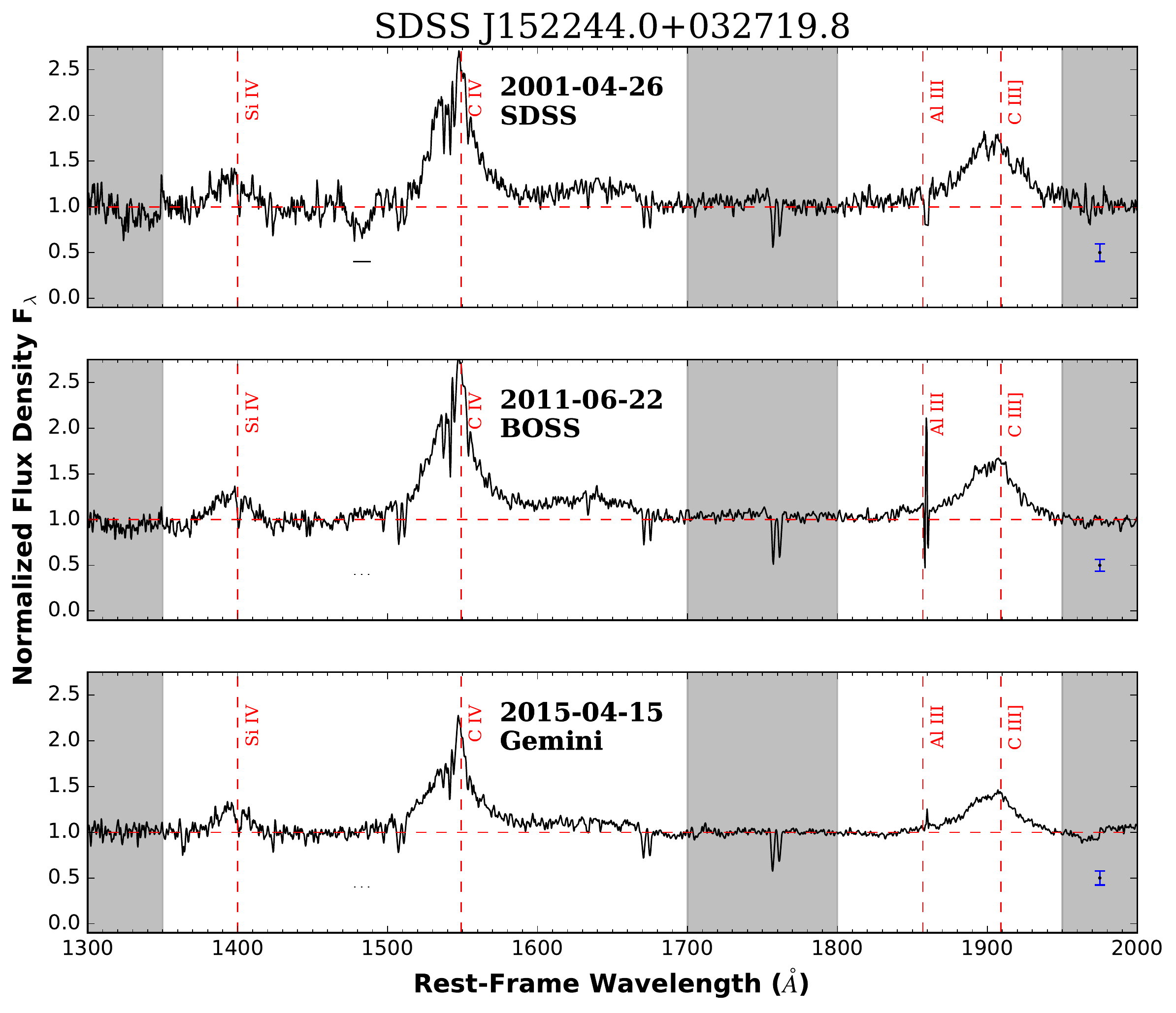}
\caption{
The normalized optical spectra of a representative transforming BAL quasar. The gray shaded areas indicate the RLF regions used to obtain the continuum fit. The horizontal black solid line extends over the range of the \civ~BAL region in the first SDSS spectrum; the BAL feature is not seen in the subsequent BOSS and \gemini~spectra as indicated by the corresponding black horizontal dotted lines. The median error bar on the normalized flux density is shown at the bottom-right of each spectrum. The complete figure set (8 images showing 28 spectra in total) for all the transforming BAL quasars is available online.
\label{Figure3}
}
\end{center}

\end{figure*}

Preceding studies on BAL variability have modeled the underlying continuum using a variety of models such as power-law fits, reddened power-law fits, and polynomial fits. Following \citet[G09]{gibson2009catalog}, we have employed a reddened power-law to define the underlying continuum. \citet{Hopkins04} reported that the intrinsic dust reddening in quasar spectra can be better modeled by Small-Magellanic-Cloud-like (SMC-like) reddening compared to Large Magellanic Cloud or Milky Way models. The continuum is thus reconstructed with an intrinsically reddened power-law model using the SMC-like reddening from \citet{Pei92}. This model is defined by the three parameters of power-law normalization, power-law spectral index, and intrinsic-absorption coefficient.

Following the technique of G09, six relatively line-free (hereafter, RLF) windows were used to fit a continuum model, subject to the condition that the spectrum has the relevant spectral coverage. The ``canonical'' RLF regions were \hbox{1250--1350}\,\AA, \hbox{1700--1800}\,\AA, \hbox{1950--2200}\,\AA, \hbox{2650--2710}\,\AA, \hbox{2950--3700}\,\AA, and \hbox{3950--4050}\,\AA. To account for the span of line-free regions that in certain cases are of significantly differing lengths, each pixel was weighted to ensure that each of the RLFs contributed equally to the continuum fit. The continuum function was iteratively fitted to the RLFs using the Levenberg-Marquardt least-squares minimization procedure \citep{markwardt2009non} which involves rejecting data points that deviate by more than 3\,$\sigma$ from the fit within the RLF windows. These steps were redone until the fit parameters converged. We present two spectra (see Fig.~\ref{Figure2}) with the continuum fits overlaid. The fit quality in the RLF regions is estimated using the reduced chi-squared statistic, $\chi^{2}_\nu$. For the evaluation of $\chi^{2}_\nu$, we binned the spectra by the corresponding resolution element to avoid the contribution of correlated noise. Each spectral epoch was divided by its corresponding best continuum fit to obtain the normalized spectrum which is then used to obtain the measurements of BAL properties as discussed in Section~\ref{sec:MeasBALproperties}. The normalized spectra for one representative transforming BAL quasar are shown in Fig.~\ref{Figure3}. Each spectrum was visually inspected to check if the fit appears appropriate in the RLF regions, and in the region between \hbox{1300--1900} \angstrom~where \civ, \siv, and \al~BALs are expected to be detected. If a fit was deemed inappropriate, additional processing was done to obtain acceptable fits. These steps included removing the sigma-clipping algorithm to prevent useful RLF windows from being ``clipped'' away, using a subset of the RLF windows, using smaller intervals within the RLF windows, and including additional RLFs of \hbox{1350--1380}\,\AA~and/or \hbox{1410--1440}\,\AA~that appeared free of absorption (these windows are devoid of strong emission and absorption in general and are a good representation of the underlying continuum). Given that we do not associate continuum-fit parameters with any physical meaning, the use of these alternative steps does not result in loss of information for further analysis. Such additional processing was needed for 7 of the 28 spectra presented in Fig.~\ref{Figure3}.

We estimated the uncertainties in the continuum fits using ``flux randomization'' Monte Carlo simulations (e.g. \citealt{peterson1998optical}) so as to vary the flux in each pixel of the spectrum by a random Gaussian deviate based on the spectral uncertainty. The continuum was then fit to the new spectrum and the process repeated 100 times. The standard deviation of the 100 iterations was adopted as the uncertainty of the continuum fit and used in the evaluation of uncertainty on the EW (see Equation~\ref{deltaEW}).

\subsubsection{Measurements of BAL properties} \label{measur}
\label{sec:MeasBALproperties}

Following the sequence of steps in the preceding section, each normalized spectrum was smoothed using the Savitzky-Golay (SG) algorithm~\citep{savitzky1964smoothing}. The SG parameters were chosen to perform local linear regression on three consecutive data points to remove artifacts originating from noise and preserve the trends of slow variations. Such smoothed spectra are used to detect the presence (or lack) of~\civ~BALs.

Each smoothed spectrum was searched for absorption troughs reaching a level of $\geq$\,10 per cent below the continuum, following the conventional definition of BAL and mini-BAL troughs \citep{wey91}. The trough width is determined by the velocities corresponding to the shortest and the longest wavelengths for a given trough, $v_{\rm max}$ and $v_{\rm min}$, respectively. The detected features were sorted into mini-BAL troughs (500--2000 $\mathrm{km\,s^{-1}}$ wide) and BAL troughs ($\geq~$2000 $\mathrm{km\,s^{-1}}$ wide). We measured the rest-frame BI, EW, and EW uncertainties from the unsmoothed data. The C\,{\sc iv} BI of each spectrum is calculated using the BI definition of \citet{wey91}:

\begin{equation}
\label{eqn:BI}
\mbox{BI} \equiv \int_{-3000}^{-25000} \left( {1-\frac{f(v)} {0.9}} \right) Cdv
\end{equation}

\noindent The units of BI are $\mathrm{km\,s^{-1}}$, where $f(v)$ is normalized flux density as a function of velocity and $C$ takes a value of 1.0 only where a trough is wider than 2000\,$\mathrm{km\,s^{-1}}$, and is 0.0 otherwise. 

In addition, the rest-frame equivalent width (EW) for each BAL trough in 
units of \AA\ is calculated; note that \hbox{1\,\AA\\} 
corresponds to \hbox{$\approx $\,200\,$\mathrm{km\,s^{-1}}$} in the 
C\,{\sc iv} absorption region. The EW was measured following:

\begin{equation}
\textrm{EW} = \sum_i \left( 1 - \frac{F_i}{F_c} \right) W_i
\label{EW}
\end{equation}
and the uncertainty on the EW
\begin{equation}
\label{deltaEW}
\sigma_{\textrm{EW}} = \sqrt{\left[\sum_i\frac{\Delta F_{ci}}{F_{ci}} \left(\frac{W_iF_i}{F_{ci}}\right) \right]^2 + \sum_i\left(\frac{W_i\Delta F_i}{F_{ci}}\right)^2}
\end{equation}

\noindent $W_i$ is the bin
width in units of \AA, $F_i$ and $\Delta F_i$ denote the flux and the error on the flux in the $i^{\rm th}$ bin, $F_{ci}$ is the underlying
continuum flux in the $i^{\rm th}$ bin, and $\Delta F_{ci}$ is the uncertainty in the continuum flux.

We tested for agreement of the BI, EW, and velocity width values estimated in the current study with those of \citet{ak2012} for SDSS I/II/III spectra by calculating the mean difference and the RMS scatter in the differences. The average differences of the BI, EW, and velocity width values are less than three times the RMS scatter, indicating there is no evidence for a systematic deviation between the methods. The reasons for slight variations between our values and those from \citet{ak2012} could be differences in continuum fitting (e.g. RLF window selection, pixel weighting, etc.). We provide the 90 per cent confidence upper limits on the EW for the BAL troughs that have disappeared. The upper limits are estimated assuming a multiplicative model based on the observed profile of the BAL trough before it disappeared using the method described in Section 15.6 of \citet{press07}.

We find that the \civ~BAL troughs that had previously disappeared remained absent in the latest observations, and no other BALs emerged. Our objects have thus remained normal non-BAL quasars (refer to Table~\ref{tab:table-1}).

\subsubsection{Notes on individual objects}
\label{sec:notesonobj}

\textit{J074650.59+182028.7}: the SDSS, BOSS, and \gemini~spectra contain Narrow Absorption Line (NAL; velocity widths $\leq$\,500~km~s$^{-1}$) doublets from \civ~that are blended with the disappearing \civ~BAL trough.

\textit{J085904.59+042647.8}: spectra contain a narrow \civ~system that is blended with the disappearing \civ~BAL trough.

\textit{J093620.52+004649.2}: spectra show the emergence of a \civ~BAL trough within a period of $\lesssim$\,105 days in the rest frame between the two SDSS observations. The BOSS spectrum revealed that this same trough has disappeared over a period of $\lesssim$\,3.3 rest-frame years. From our contemporaneous \hbox{\gemini}~observations, we observe that the BAL trough has remained absent for at least 1.3 rest-frame years. 

\textit{J114546.22+032251.9}: spectra show the presence of detached \civ~NAL doublets not overlapping with the disappearing \civ~BAL trough in all the epochs of observation.

\textit{J133211.2+392825.9}: the SDSS spectrum of J133211.2+392825.9 shows the presence of a weak absorption component at 1504.6--1510.1~\angstrom~in the rest frame that does not satisfy the \hbox{2000~km~s$^{-1}$} minimum-width criterion to be considered as a BAL.

\section{RESULTS}
\label{sec:Results_Discussion}

\noindent The optical/UV spectral analysis demonstrates that the BAL troughs have remained absent in the new \hbox{\gemini}~and ARC 3.5-m observations, indicating that material capable of producing such UV absorption is generally absent from our line-of-sight. Our \hbox{\chandra}~observations can test if the \hbox{X-ray} characteristics also show similar behavior; i.e., a lack/weakening of \hbox{X-ray} absorption. We investigate this question by comparing the distributions of \daox~for different BAL populations (Section~\ref{sec:survanalysis}), by carrying out joint spectral analysis (Section~\ref{sec:jointspectralanalysis}), and by constructing the Spectral Energy Distributions (SEDs) of our transforming BAL quasars and comparing them with BAL and non-BAL populations (Section~\ref{sec:SED}).

\subsection{Distributions of \aox~and \daox}
\label{sec:survanalysis}

\begin{figure*}
\centering
\includegraphics[width=\textwidth]{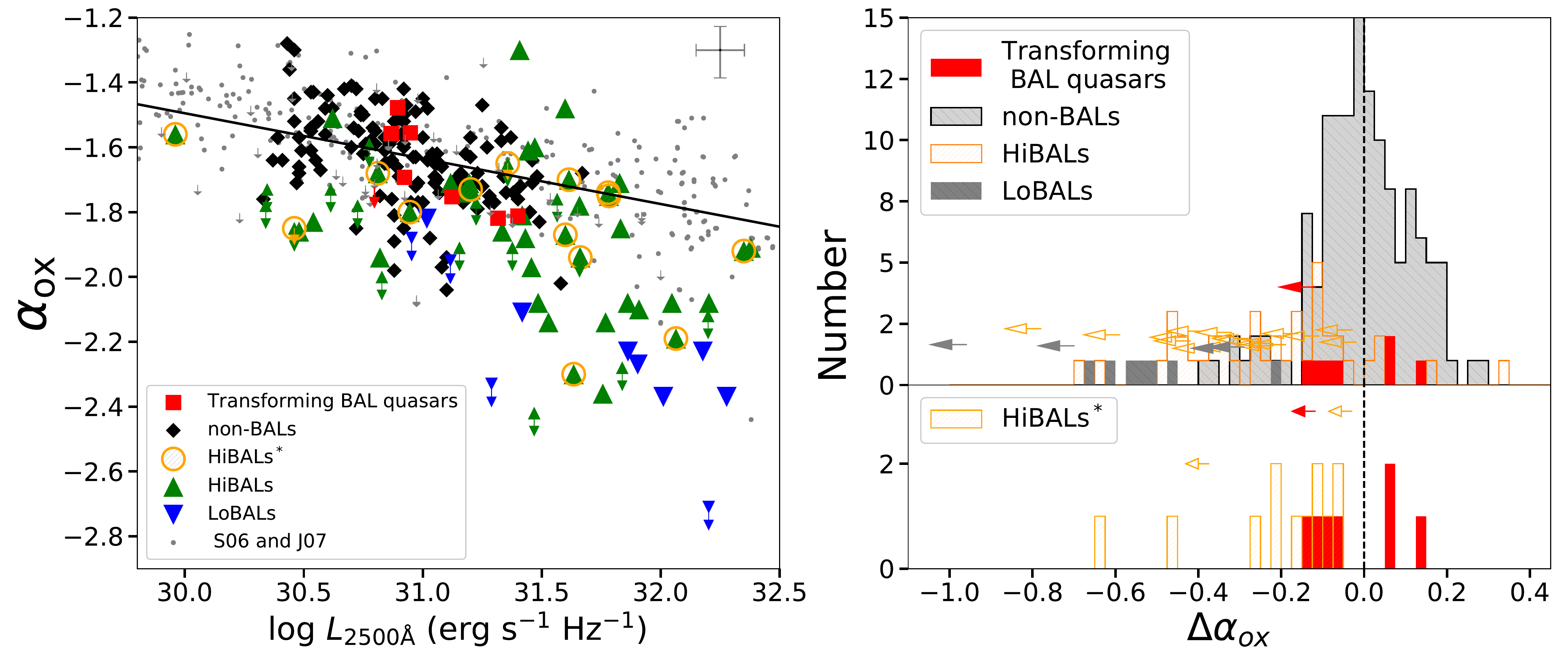} 
\caption{Left panel: \hbox{X-ray}-to-optical power-law slope ($\alpha_{\rm ox}$) vs. monochromatic luminosity at rest frame 2500~\AA~for our transforming BAL quasar sample (squares and a downward arrow). Other samples in the figure are non-BALs from G08, shown as black dots, a combination of AGN samples from S06 and 
J07 shown as gray dots and gray downward arrows, and those of HiBALs, HiBALs$^{*}$, and LoBALs taken from G09, F08, and S11. The solid line represents the best-fit relation between 
$\alpha_{\rm ox}$ and $L_{2500\mbox{\rm~\scriptsize\AA}}$ from J07. The median uncertainties of $\alpha_{\rm ox}$ and $L_{2500\mbox{\rm~\scriptsize\AA}}$ for our transforming BAL quasars are shown as gray error bars in the top-right of the panel. 
Right panel: Distribution of $\Delta \alpha_{\rm ox}$ for our transforming BAL-quasar sample (red-shaded histogram and red leftward arrow) compared with that of the 132 non-BALs from G08, and those of optically selected HiBALs, HiBALs$^{*}$, and LoBALs taken from G09, F08, and S11. The vertical dashed line indicates $\Delta \alpha_{\rm ox} = 0$.}
\label{Figure4}
\end{figure*}

Previous \hbox{X-ray} studies of absorbed BAL quasars (e.g. \citealt{green96,gallagher2006exploratory,gibson2009catalog}) have used \hbox{X-ray}-to-optical spectral slopes to assess the \hbox{X-ray} continuum level. Several studies (e.g. \citealt{brandt2000nature,gallagher2006exploratory,wu2010x}) have reported correlations of properties such as BI and \civ~EW with \aox~and \daox. 

Fig.~\ref{Figure4} displays a comparison of \aox~and \daox~for different BAL quasars. For comparison, a combination of AGN samples from \citet[S06]{steffen2006x} and J07 are shown; a strong anti-correlation of \aox~with $L_{2500\mbox{\rm~\scriptsize\AA}}$~is visible, as indicated by the black line in Fig.~\ref{Figure4}. We therefore used \daox~to compare the different BAL quasar types in order to remove the long-established luminosity dependence of \aox; note that a quasar with given \daox~is \hbox{X-ray} weak compared to the average quasar by a factor of $f_{\text{weak}}$ $\approx$ 403$^{-\Delta\alpha_{\rm ox}}$. We compared the \daox~distribution of our transforming BAL quasars with those of optically selected HiBAL quasars (42 in number taken from G09, 10 from \citealt[][F08]{fan2008correlation}, and 6 from \citealt[][S11]{stalin2011x}), LoBAL quasars (8 in number also taken from G09, 3 from F08) and non-BAL quasars (132 in number taken from \citealt[][G08]{gibson2008optically}). For one of the objects from S11, we derived the $f_\text{2 keV}$ and subsequently \daox~from \citet{chen2018xmm}, as this source only had an \hbox{X-ray} upper limit in the shallower \hbox{X-ray} data used by S11. We also constructed a subset-HiBAL population based on the BI values for our transforming BAL quasars in their earliest epoch of observation. The BI values for our transforming BAL quasars have a median value of \hbox{803 km s$^{-1}$} and a range of \hbox{309--1260~km~s$^{-1}$} (refer to Table~\ref{tab:table-1}). We chose HiBAL quasars within this same matched BI range from G09 (10 HiBALs), F08 (2 HiBALs), and S11 (2 HiBALs) and compared the \daox~distribution of our transforming BAL quasars with those of the 14 chosen HiBAL quasars (we define these objects as HiBALs$^{*}$).

The \daox~parameter in the case of our target J085904.59+042647.8 is an upper limit; similarly, some of the quasars in our comparison sets have \daox~upper limits. Hence, we apply statistical methods that can recover information from censored data (e.g. \citealt{feigelson2012modern}). The \textit{survfit} function in R was used to compare the \daox~distributions with the modified Peto-Peto test~\citep{peto1972asymptotically}. When comparing the transforming BAL quasars and non-BALs, the $p$-value is 0.630\,$\pm$\,0.006, indicating that the difference between the two sets is not statistically significant.\footnote{\label{footnote:boot} We estimated the standard error of the mean for the $p$-values using the method of nonparametric bootstrapping~\citep{babu1983inference} in order to determine how the $p$-value depends on plausible random variations in the observations by carrying out random sampling from within the dataset using 1000 replicates.} We find the transforming BALs have a statistically significant difference from HiBALs and LoBALs with $p$-values of 0.040\,$\pm$\,0.004 and 0.0015\,$\pm$\,0.0001, respectively. Furthermore, our transforming BAL quasars show a statistically significant difference from the HiBALs$^{*}$ with a $p$-value of 0.039\,$\pm$\,0.004. Thus, transforming BAL quasars do not appear to arise from the same population as HiBALs$^{*}$, HiBALs and LoBALs. 

We also calculated the mean values of \daox~for the different BAL-quasar types discussed above using the Kaplan-Meier estimator (implemented using the \textit{survfit} function in R). The transforming BAL quasars have a mean \daox~value of $-$0.043\,$\pm$\,0.028, which is consistent with the mean \daox~value of $-$0.007\,$\pm$\,0.011 for non-BAL quasars. In comparison, the mean \daox~values for HiBALs$^{*}$, HiBALs, and LoBALs are $-$0.238\,$\pm$\,0.050, $-$0.303\,$\pm$\,0.028, and $-$0.587\,$\pm$\,0.033, respectively. 

\subsection{Joint spectral fitting of transforming BAL quasars}
\label{sec:jointspectralanalysis}

The objects in our study have a total of 94 counts, and individually they have too limited numbers of \hbox{X-ray} photons to enable reliable individual spectral fitting. In order to set a useful constraint on the average power-law photon index of these eight sources as an ensemble, joint spectral fitting was performed. Before performing this group analysis, the spectrum of each individual source and its background was extracted using the {\hbox{\sc CIAO SPECEXTRACT}} script within a circular aperture of 4\arcsec~radius centered on the \hbox{X-ray} position. The background spectrum was extracted from an annular region free of \hbox{X-ray} source contamination with inner radius 10\arcsec\ and outer radius 40\arcsec. Joint spectral analysis of the eight transforming BAL quasars was performed with XSPEC (version 12.10.0; \citealt{arnaud96}) using the \hbox{$C$-statistic} \citep{cash1979parameter} which is appropriate to the low-count scenario of our analysis (e.g. \citealt{nousek1989chi}). The \hbox{X-ray} spectrum for each source was sampled in order to have at least one count per energy bin. This procedure was carried out to avoid zero-count bins while permitting retention of all spectral information. The fitting was done with a simple model consisting of a power law and Galactic absorption, permitting each source to be assigned its own Galactic column density. The quoted errors are at the 68 per cent confidence level for one parameter of interest. The best-fit photon index is \hbox{$\Gamma$\,=\,$1.69^{+0.25}_{-0.25}$} ($C$-statistic\,=\,70.04 using 85 bins), which is consistent with the typical photon index \hbox{$\Gamma\,\approx$\,1.8--2.0} for radio-quiet quasars (e.g. \citealt{reeves1997x,page2005xmm}; J07; \citealt{shemmer2008hard,scott2011new}).\footnote{We also checked if our fit is strongly affected by the brightest sources in our sample. For example, we find the best-fit photon index to be \hbox{$\Gamma$\,=\,$1.65^{+0.32}_{-0.32}$} with the brightest object (SDSS J$133211.21+392825.9$) removed. Although the uncertainties become larger, the best-fit photon index remains consistent with those of typical radio-quiet quasars.} BAL quasars, on the other hand, are known to be \hbox{X-ray} absorbed and are generally found to exhibit harder \hbox{X-ray} spectra due to the paucity of low-energy \hbox{X-ray}s, leading to smaller values of effective photon index. Their effective photon indices are typically \hbox{$\Gamma$\,=\,0--1.4}, with a mean value of $\Gamma\,\approx$\,1. For example, if we stack the 35 LBQS BAL quasars in \citet{gallagher2006exploratory}, we derive an average \hbox{$\Gamma$\,=\,$1.05^{+0.22}_{-0.15}$} which is inconsistent with the mean value for our transforming BAL quasars of \hbox{$\Gamma$\,=\,$1.69^{+0.25}_{-0.25}$}.

We also joint fit our spectral data with a model incorporating neutral intrinsic absorption adopting the individual redshift of each source. This model returned an intrinsic column density consistent with zero \hbox{($N_{H}$\,$\textless$\,0.77~$\times$ 10$^{22}$~cm$^{-2}$, 90 per cent confidence upper limit}). Typically, BAL quasars show stronger \hbox{X-ray} absorption than this upper limit with column densities of order \hbox{$N_{H}$\,$\sim$\,10$^{23}$~cm$^{-2}$}~(e.g. \citealt{green96,mathur2000,green01}; \citealt[][]{gallagher2002x,gallagher2006exploratory}; \citealt{gruped}).

\subsection{IR-to-\hbox{X-ray} spectral energy distributions}
\label{sec:SED}

\begin{figure}
\centerline{
\includegraphics[width=0.50\textwidth]{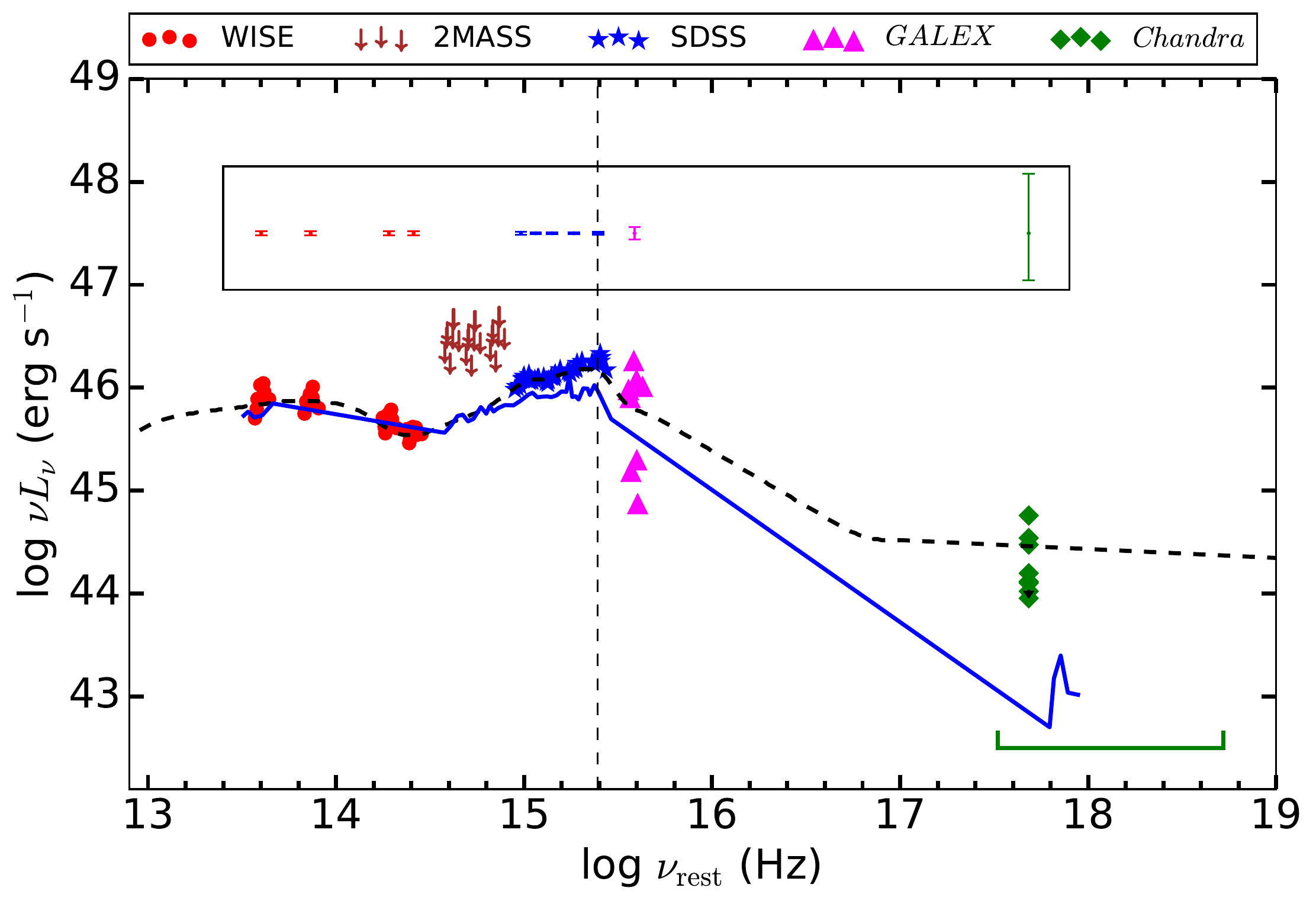}
}
\caption{
Combined SEDs of the eight transforming BAL quasars. The IR-to-\hbox{X-ray} SED data are from {\it WISE} (dots), 
2MASS (downward arrow), SDSS (stars), {\it GALEX} (triangle), and \hbox{\chandra}\ (diamond). The median errors in each band are presented inset. The horizontal segment represents the \hbox{\chandra}\ bandpass in the rest frame at the average redshift \hbox{($z$\,=\,1.94)} of our targets.
Galactic extinction has been accounted for in each band. All of our targets have been normalized to the composite SED (dashed line) of optically luminous quasars 
\citep{richards2006spectral} at \hbox{3000~\AA} (corresponding to $10^{15}$~Hz). The vertical dashed line at $2.47~\times 10^{15}$~Hz corresponds to the Ly$\alpha$ line. 
The thick solid line represents the mean SED of BAL quasars from \citet{gallagher2007radio}.
All of our targets are comparable in \hbox{X-ray} intensity to the mean SED of normal quasars and show significant excess \hbox{X-ray} emission relative to the mean SED of BAL quasars.
}
\label{Figure5}
\end{figure}

Normalized SEDs in the infrared-to-\xray\ regime for our targets are presented in Fig.~\ref{Figure5}. The data were collected from the Wide-field Infrared Survey Explorer ({\it WISE\/}), Two Micron All Sky Survey (2MASS), SDSS, 
{\it Galaxy Evolution Explorer} ({\it GALEX\/}), and \hbox{\chandra}.

The near-infrared portion of the SEDs were constructed using the ${\it WISE\/}$ observations in four bands centered at 3.4, 4.6, 12, and 22~$\mu$m. All of our transforming BAL quasars have detections in these bands. The flux densities of our targets were calculated from {\it WISE\/} Vega magnitudes with the application of color corrections \citep{wright2010wide}. 

All of our eight targets were undetected in the 2MASS \hbox{near-infrared} bands: $J$ (1.25~$\mu$m), 
$H$ (1.65~$\mu$m), and $K_{\rm s}$ (2.16~$\mu$m). For these targets, a 2MASS sensitivity, \hbox{S/N\,=\,3}, was used to estimate the flux-density limit in each of the bands. For the {\it GALEX\/} data, we have used only the NUV data (not including any FUV data), and all of our targets have detections in the {\it GALEX\/} NUV band; however, the data points from the {\it GALEX} NUV band and the SDSS $u$-band for quasars with redshifts above $\approx$ 0.86 and 1.91 are affected by Ly$\alpha$ forest absorption which introduces additional systematic uncertainty in their SED values because of its presence at wavelengths shortward of rest-frame \hbox{1216~\AA}. The \xray\ and optical data analysis have been discussed earlier in Sections~\ref{sec:chandraanalysis} and~\ref{sec:opticalanalysis}, respectively. We compare the mean SED of optically luminous SDSS quasars from \citet{richards2006spectral} with those of transforming BAL quasars in Fig.~\ref{Figure5}. The SEDs of our targets have been scaled to the \citet{richards2006spectral} 
mean SED at rest frame 3000~\AA\ (corresponding to a frequency of $10^{15}$~Hz), and they are broadly in agreement in the infrared-to-UV regime with the mean SED for quasars. 

We have also compared the SEDs of our transforming BAL quasars with the typical SED of BAL quasars. The blue solid line in Fig.~\ref{Figure5} represents the mean SED of BAL quasars from \citet{gallagher2007radio}. The flux densities at 2~keV closely resemble the SED of normal quasars and are relatively higher than the mean SED of BAL quasars, providing confidence that our transforming BAL-quasars are similar to non-BAL quasars. This SED-based approach thus vindicates our use of \aox~and \daox~in Section~\ref{sec:survanalysis}.

\section{SUMMARY, DISCUSSION, AND FUTURE WORK}
\label{sec:summary}
\noindent In this paper, we have presented \hbox{X-ray} and multiwavelength investigations of the nature of a sample of eight SDSS BAL quasars that transformed into non-BAL quasars. The key points of this work are the following:

\begin{enumerate}  

\item We obtained \hbox{\chandra}~exploratory (2.5--4.9~ks) observations of all eight transforming BAL quasars. We measured their basic \hbox{X-ray}, optical, and radio properties. See Section~\ref{sec:chandraanalysis} and Table~\ref{tab:table-2}.

\item We acquired new optical/UV spectral epochs from the \hbox{\gemini}~and ARC 3.5-m telescopes to provide contemporaneous constraints upon rest-frame UV BAL absorption and continuum emission, aiding the interpretation of our \hbox{\chandra}~data. These spectral data reveal that the BAL troughs that disappeared remain absent in all quasars. See Section~\ref{sec:opticalanalysis}.

\item We compared the \aox~and \daox~distributions of our transforming BAL quasars with those for samples of BAL and non-BAL quasars to determine which they resemble, using statistical tests. Our analysis shows that the transforming BAL quasars are akin to non-BAL quasars. See Section~\ref{sec:survanalysis}. 

\item We performed joint \hbox{X-ray} spectral fitting in order to set a constraint on the effective power-law photon index of these sources as an ensemble and obtained a best-fit photon index of \hbox{$\Gamma$\,=\,$1.69^{+0.25}_{-0.25}$}. This value is consistent with that expected for typical radio-quiet quasars and is steeper than expected for most BAL quasars. We also find no evidence for intrinsic \hbox{X-ray} absorption on average. See Section~\ref{sec:jointspectralanalysis}.

\item We constructed IR-to-\xray\ continuum SEDs for the transforming BAL quasars. The transforming BAL quasars are similar to non-BAL quasars in their overall SED continuum properties, but not to BAL quasars. See Section~\ref{sec:SED}.

\end{enumerate}

\noindent Overall, our \hbox{\chandra}\ SED and spectral results indicate that 
any \xray\ absorbing material largely moved out of our line-of-sight 
when these quasars transformed from BAL to non-BAL quasars. 
In the context of the shielding-gas model (see Section~\ref{sec:Intro}), such 
a reduction in the amount of \xray\ absorption could arise due 
to accretion-disk rotation or physical restructuring of the
shielding material; this change would allow increased \hbox{X-ray} and EUV radiation to reach the wind material producing the UV BALs. A sufficient corresponding increase in the ionization level 
of this material could lead to the observed disappearance of 
the UV BALs, owing to reductions in the relevant ion column 
densities. This scenario of changes in the amount of 
shielding also naturally explains the coordinated variations 
of BAL-trough strengths established even for distinct \civ~BAL troughs separated by thousands of km~s$^{-1}$ (e.g. \citealt{ak2012,ak2013broad,decicco2018}). Absorption components from \civ~that are widely separated in velocity, and thus arise in different locations along the line of sight, could still all respond in a coordinated manner to changes in the incident \hbox{X-ray}/EUV flux. 

Our \hbox{\chandra}\ results can also be interpreted in the context of models that invoke wind clumping, rather than shielding gas, to prevent overionization of the material producing the 
UV BALs (see Section~\ref{sec:Intro}). In this scenario, the \hbox{X-ray} absorption and UV BALs arise in the same, or closely related, material, so the observed coordinated variations of the \hbox{X-ray} absorption and UV BALs are expected. Current clumping-based models do not provide an obvious mechanism for explaining the observed coordinated variations of distinct \civ~BAL troughs widely separated in velocity and location, but further development of these models may provide a mechanism. 

There are several avenues by which our work might be advanced observationally. Since the investigation of 
\citet{ak2012} that identified the transforming 
BAL quasars studied here, additional examples of
such objects have been found. In particular, \citet{decicco2018} have recently identified 30 transforming BAL quasars
in their systematic analysis of a large SDSS quasar sample, 
and these objects could be efficiently targeted in \hbox{X-ray}s to improve our sample size. Furthermore, the objects studied in this paper should continue to be monitored with 
optical/UV spectroscopy to search for any re-emergence of 
the BALs that disappeared; examples of such re-emergence 
have been discovered in other BAL quasars 
(e.g. \citealt{mcgraw2017broad,rogerson2018}). 
If such re-emergence events are found, then prompt \hbox{X-ray} 
observations could search for a corresponding increase of 
\hbox{X-ray} absorption, thereby further examining the connection
between \hbox{X-ray} absorption and UV BAL variability. Higher cadence optical/UV spectroscopy of BAL-disappearance events could also place novel constraints upon the nature of this variability and the radial location of the BAL material.

\section*{ACKNOWLEDGEMENTS}
We thank the anonymous referee for his/her suggestions which helped improve the manuscript. We acknowledge helpful discussions with Chien-Ting Chen, Eric Feigelson, Sean McGraw, Qingling Ni, Yue Shen, Fabio Vito, Guang Yang, Weimin Yi, Ning-Xiao Zhang, and Shifu Zhu. S and WNB acknowledge support from NSF grant AST-1516784 and Chandra \hbox{X-ray} Center grant GO5-16089X. NFA acknowledges support from TUBITAK 115F037. BL acknowledges financial support from the National Key R\&D Program of China grant 2016YFA0400702 and National Natural Science Foundation of China grant 11673010. 

This work contains data based on observations obtained at the Gemini Observatory, which is operated by the Association of Universities for Research in Astronomy, Inc., under a cooperative agreement with the NSF on behalf of the Gemini partnership: the National Science Foundation (United States), the National Research Council (Canada), CONICYT (Chile), Ministerio de Ciencia, Tecnolog\'{i}a e Innovaci\'{o}n Productiva (Argentina), and Minist\'{e}rio da Ci\^{e}ncia, Tecnologia e Inova\c{c}\~{a}o (Brazil). This paper also made use of data based on observations obtained with the Apache Point Observatory 3.5-meter telescope, which is owned and operated by the Astrophysical Research Consortium.

\bibliographystyle{mnras}
\bibliography{references}

\bsp
\end{document}